\documentclass[a4paper,fleqn,usenatbib]{mnras}
\hypersetup{final=false}

\usepackage{mathptmx}

\usepackage[T1]{fontenc}
\usepackage{ae,aecompl}

\usepackage{graphicx}	
\usepackage{amsmath}	
\usepackage{amssymb}	

\usepackage{color}
\newcommand{\rev}[1]{#1}

\title[Spiral arms in protoplanetary discs]{Spiral arms in thermally stratified protoplanetary discs}

\author[A. Juh\'{a}sz and G. P. Rosotti]{Attila Juh\'{a}sz$^{1}$\thanks{juhasz@ast.cam.ac.uk} and Giovanni P. Rosotti$^{1}$\thanks{rosotti@ast.cam.ac.uk}\\
$^{1}$ Institute of Astronomy, Madingley Road, Cambridge CB3 OHA, UK\\
}

\date{Accepted XXX. Received YYY; in original form ZZZ}

\pubyear{2017}

\begin{document}
\label{firstpage}
\pagerange{\pageref{firstpage}--\pageref{lastpage}}
\maketitle

\begin{abstract}
Spiral arms have been observed in nearly a dozen protoplanetary discs in near-infrared scattered light and recently also in the sub-millimetre continuum. While one of the most compelling explanations is that they are driven by \rev{planetary or stellar companions, in all but one cases such companions have not yet} been detected and there is even ambiguity on whether the planet should be located \textit{inside} or \textit{outside} the spirals. Here we use 3D hydrodynamic simulations to study the morphology of spiral density waves launched by embedded planets taking into account the vertical temperature gradient, a natural consequence of stellar irradiation. Our simulations show that the pitch angle of the spirals in thermally stratified discs is the lowest in the disc mid-plane and increases towards the disc surface. We combine the hydrodynamic simulations 
with 3D radiative transfer calculations to predict that the pitch-angle of planetary spirals observed in the near-infrared is higher than in the sub-millimetre. We also find that in both cases the spirals converge towards the planet. This provides a new powerful observational method to determine if the perturbing planet is inside our outside the spirals, as well as map the thermal stratification of the disc.
\end{abstract}




\section{Introduction}
\label{sec:introduction}
With the dawn of extreme high resolution imaging capabilities of current telescopes (e.g. SPHERE/VLT, GPI/Gemini, ALMA) more and more discs show spiral arms in near-infrared (NIR) scattered light (e.g. \citealt{muto_2012,benisty_2015,wagner_2015,stolker_2016,benisty_2017}) and
 recently also in the sub-millimetre \citep{perez_2016,tobin_2016}. The observed spirals have similar morphologies in most cases; two symmetric arms,  shifted in azimuth by approximately 180\degr, and with a pitch angle of about 15\degr--30\degr. In most cases there is also a gap / cavity just inwards of  the spirals.
 
It has been known for a long time that spiral density waves are launched by \rev{planetary or stellar companions} at Lindblad-resonances \citep[e.g.,][]{goldreich_1979, ogilvie_2002, rafikov_2002a} both inwards and outwards of the \rev{companion's} orbit. 
Therefore a straightforward explanation for the observed spirals is that they are launched by so far undetected 
\rev{companion (possibly a forming, embedded planet)}, (e.g. \citealt{muto_2012,grady_2013}); this would open up the exciting possibility of getting a direct insight into planet formation. Using 2D hydrodynamic simulations and 3D radiative transfer computation \citet{juhasz_2015} showed, however, that the pitch angle and contrast of the spirals driven by planets orbiting \emph{inwards} of the spirals are not compatible with the NIR observations. \citet{dong_2015a} combined 3D hydrodynamic simulation with 3D radiative transfer calculations to show that the observed contrast and pitch angle of spirals are in fact in agreement with the spirals being launched by massive (several M$_{\rm Jup}$) planets \emph{exterior} to the spirals.  

However, planets are not the only explanations for the observed spirals. Multi-armed spiral structure can also form
due to gravitational instability (GI, see e.g. \citealt{rice_2003}) with the number of arms depending on the mass and temperature of the disc via the Toomre $Q$ parameter \citep{cossins_2009}.
For very massive discs ($M_{\rm disc}/M_{\star}\gtrsim0.25$), GI can produce two armed spirals in the disc, similar to the NIR scattered light observations \citep{dong_2015b}. \citet{meru_2017} showed that GI could explain the
spirals in the sub-millimetre ALMA images of Elias\,2--27. 

Additionally, if a marginally gravitationally stable disc is perturbed by an embedded planet, the tidal interaction might tip the balance and trigger GI in the disc \citep{pohl_2015}. This combined mechanism may also produce spirals with
pitch angles similar to the ones in observations. 

Numerous theoretical studies have recently increased the level of complexity in the modelling of spirals, showing the  importance to understand their origin. For instance, while tidally induced \rev{spiral density waves}
have been studied for a long time in 2D, \citet{zhu_2015} demonstrated the \rev{spiral} amplitude and pitch angle \rev{(the complement of the angle between radial direction and the tangent of the spiral)}  in the non-linear regime are higher in 3D than in 2D simulations. However, an important aspect of a realistic disc model has not yet been investigated. Protoplanetary discs in which spirals have been detected are passively heated by the absorbed stellar radiation. As a consequence of this they will have a positive vertical temperature gradient \citep{chiang_1997}, resulting in a vertical thermal stratification in the disc. It is known that the pitch angle of the spirals depends on the local sound speed \citep{rafikov_2002a}, thus we expect vertical thermal stratification to affect the 
morphology of the spiral wake. The goal of this study is to investigate the morphology of tidally induced spirals by embedded planets in 3D \emph{thermally stratified} discs. \rev{We note, however, that our results would also be applicable to stellar companions as the tidal interaction between the disc an the companion is the same as for the planetary case.}

\section{Modeling}
\label{sec:sim_setup}

\begin{figure}
\center
\includegraphics[width=8cm]{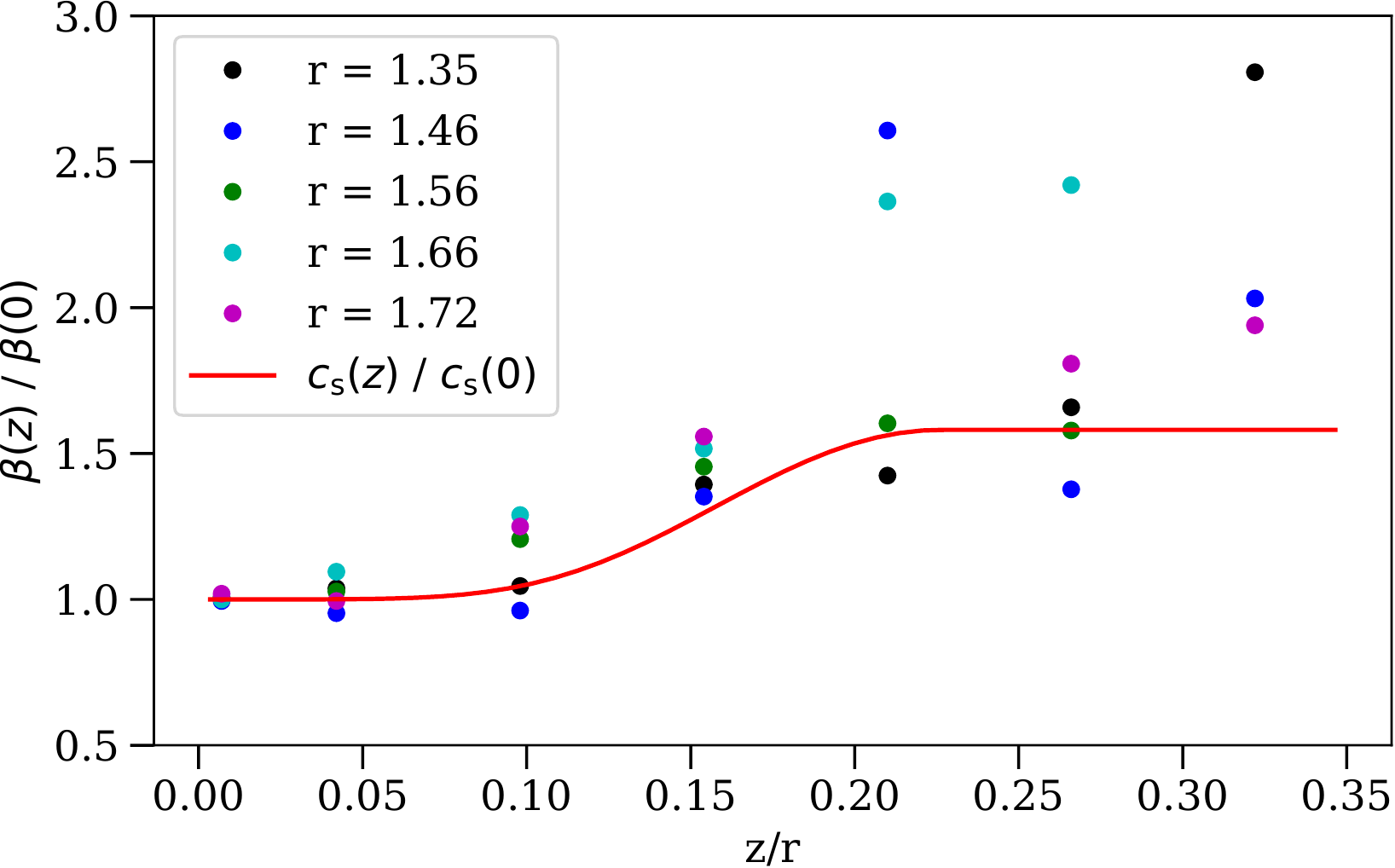}
\caption{Dependence of the spiral pitch angle on the latitude in the stratified simulation. The red line shows the sound speed profile used in the hydrodynamic simulations. The large spread in the points close to the upper boundary comes from the difficulty of determining the spiral pitch angle in that region.
}
\label{fig:pitch_angle_vertical}
\end{figure}

\subsection{Hydrodynamic simulations}
\label{subsec:hydro_setup}
We run 3D hydrodynamical simulations with the code PLUTO \citep{pluto} to investigate the disc response to the planet. We include the FARGO algorithm \citep{plutofargo} to reduce the computational cost and use the HLLC Riemann solver, linear reconstruction and second order Runge-Kutta time integration. \rev{We employ a spherical grid with 400, 50 and 
1024 cells, distributed in the [0.4, 2.5], [1.22, $\pi/2$], and [0, $2\pi$], intervals in the radial, polar and azimuthal
directions. The planet was placed at the coordinates [1.0, 0, $\pi/4$].}
In the radial and polar directions we define ``wave-killing zones" where we damp disturbances, following \citet{de-val-borro_2006}, extending up to 0.44 and from 2.25 in radius and until 1.27 in $\theta$. 
We employ a locally isothermal equation of state, in which the sound speed $c_s$ is a function of position only, \rev{and a physical viscosity using the \citet{shakurasunyaev} prescription with a value of $\alpha=10^{-3}$. We assume the planet mass to be 3\,M$_{\rm Jup}$.} 

We run two simulations, the first (subsequently called ``isothermal") in which $c_s$ depends only on radius: $c_s \propto r^{-1/4}$, with the normalization set such that the disc aspect ratio at the location of the planet is 0.07, a typical value at tens of AU. In the second simulation (``stratified"), we allow the sound speed to vary as a function of the vertical coordinate. \rev{To this end we use the prescription commonly employed when fitting observations of \citet{dartois_2003} and subsequently updated by \citet{rosenfeld_2013}. We assume that the temperature in the upper layer of the disc is 2.5 times the value in the midplane (the same as in the isothermal case), and that the transition between the cold mid-plane and the hot upper layer (the parameter $z_q$ of \citealt{dartois_2003}) happens at 3 scale-heights. } The initial surface density $\Sigma$ of the disc follows $\Sigma \propto r^{-1}$. We solve for the hydrostatic equilibrium in the vertical direction to assign the initial density at every point and we initialise the velocity so that the gas is initially in Keplerian rotation. We evolve the simulations for 100 orbits to reach a steady state in the structure of the spirals.

\subsection{Radiative transfer simulations}
\label{subsec:rad_setup}
We calculate images from the hydrodynamics simulations using the 3D radiative transfer code  
\textsc{radmc-3d}\footnote{\url{http://www.ita.uni-heidelberg.de/~dullemond/software/radmc-3d/}}, 
using the same 3D spherical mesh of the hydrodynamic simulations. We assume that the central star is a Herbig Ae
star (T$_{\rm eff}$=9500\,K, M$_\star$=2.5M$_\odot$, R$_\star$=2.0R$_\odot$), \rev{a constant gas-to-dust ratio of 100} and we calculate the dust opacity assuming astronomical silicate \citep{weingartner_2001} for a grain size distribution of $n(a)\propto a^{-3.5}$ between  {$a=0.1$\micron}  and $a=1$\,mm. With \textsc{radmc-3d} we first calculate the dust temperature with a thermal Monte-Carlo simulation and we then calculate images at wavelengths {$\lambda=1.65$\micron} and {$\lambda=1.3$\,mm} and for an inclination of 0\degr. Finally we generate synthetic observations by convolving the calculated images with a Gaussian kernel with an FWHM of {0.04\arcsec}, typical of current state-of-the-art near-infrared cameras such as SPHERE/VLT or GPI/Gemini and of ALMA \citep{alma-partnership_2015}. We assume a source distance of 100\,pc.  

\begin{figure*}
\center
\includegraphics[width=14.4cm]{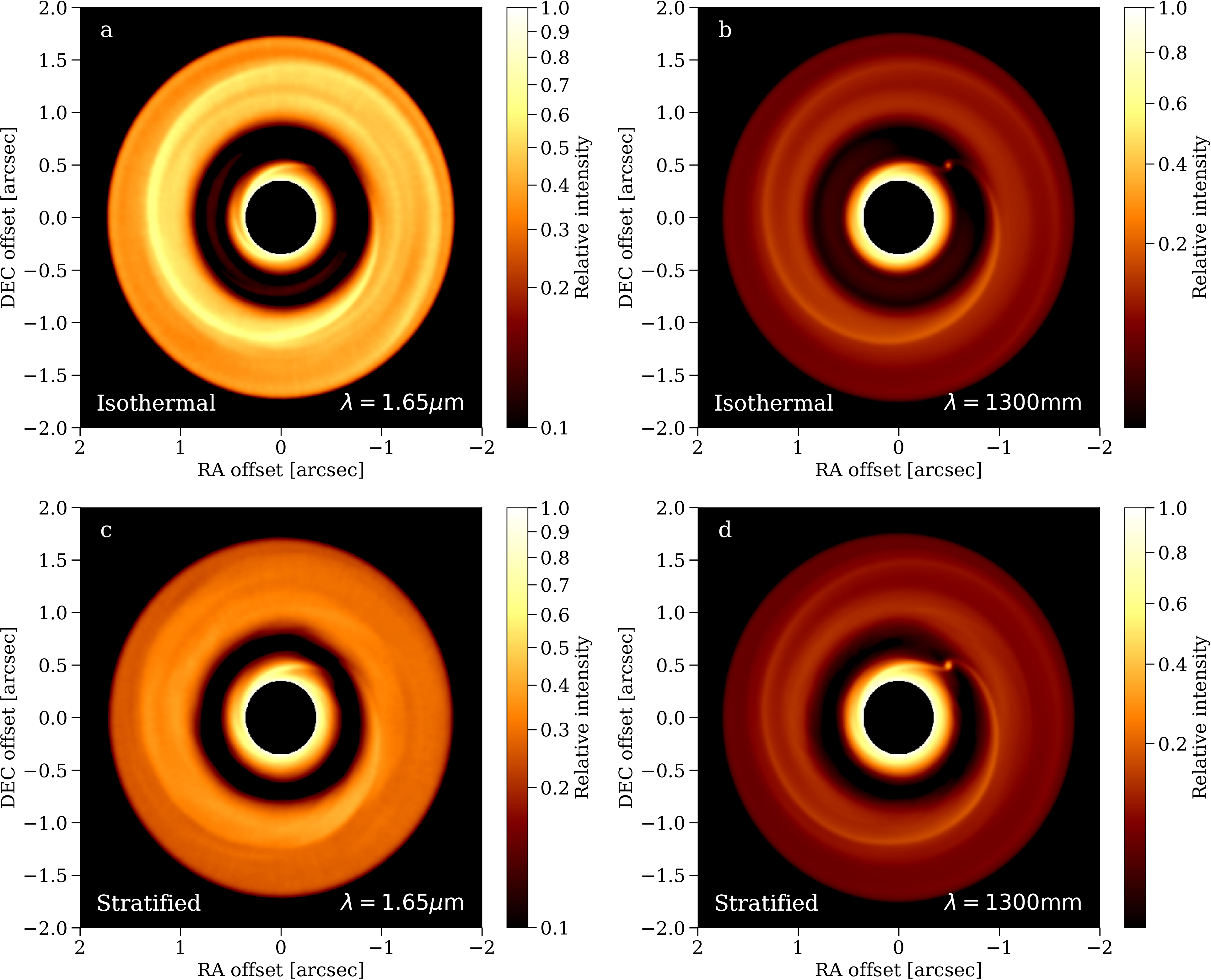}
\caption{Synthetic images in near-infrared scattered light (Panels a and c) and in sub-millimetre (Panels b and d) based on 3D hydrodynamic 
simulations assuming a vertically isothermal (Panels a and b) and stratified (Panels c and d) temperature structure. The scattered light images are 
scaled with the square of the radial distance to the star. In the near-infrared the isothermal simulation shows higher contrast spirals with lower
pitch angle compared to the stratified simulation. At sub-millimetre wavelengths the pitch angle and the amplitude of the spirals are rather similar
in the vertically stratified and isothermal simulations.}
\label{fig:images}
\end{figure*}

\section{Results}
\label{sec:results}

\subsection{Hydrodynamic simulations}

\label{sec:res_sims}

We show in figure \ref{fig:pitch_angle_vertical} the pitch angle of the spiral in the stratified simulation as a function of the height above the midplane. We plot with the colours shown in the legend the pitch angle at different radii, normalised to the value in the midplane. The red line shows how the sound speed varies with the vertical coordinate. We find that the pitch angle increases with $z$ at all radii; in addition the dependence of the pitch angle closely resembles the one of the sound speed. 

This behaviour is the natural consequence of the temperature dependence
of the pitch angle. The spirals are located where constructive interference occurs between waves emitted from Lindblad resonances in the disc \citep{ogilvie_2002}; this location depends on the wave propagation speed, and thus the disc temperature. Indeed \citet{rafikov_2002a} showed that the pitch angle of a tidally induced spiral in a 2D Keplerian disc, in which the sound speed varies with radius as
$c_s=c_p(r/r_p)^{-q}$, depends on the sound speed:
\begin{equation}
\beta = \pi/2-\tan^{-1}\left\{\frac{r_p}{h_p}\left(\frac{r}{r_p}\right)^{q+1}\left[\left(\frac{r_p}{r}\right)^{3/2} -1\right]\right\}
\end{equation}
where $\beta$ is the pitch angle, $r_p$ is the orbital radius of the planet and $h_p$ is the pressure scale height (which depends on the sound
speed via $h_p=c_p/\Omega_p$). The resemblance between the dependence with the vertical coordinate of the pitch angle and the sound speed profile may imply that this relation holds also in a 3D disc, provided that the local temperature is used in the relation. We leave a detailed analysis of the pitch angle in 3D to further studies and conclude that the vertical temperature gradient increases the pitch angle of the spirals in the upper layers of the disc.

\subsection{Simulated observations}

\begin{figure*}
\center
\includegraphics[width=14.9cm]{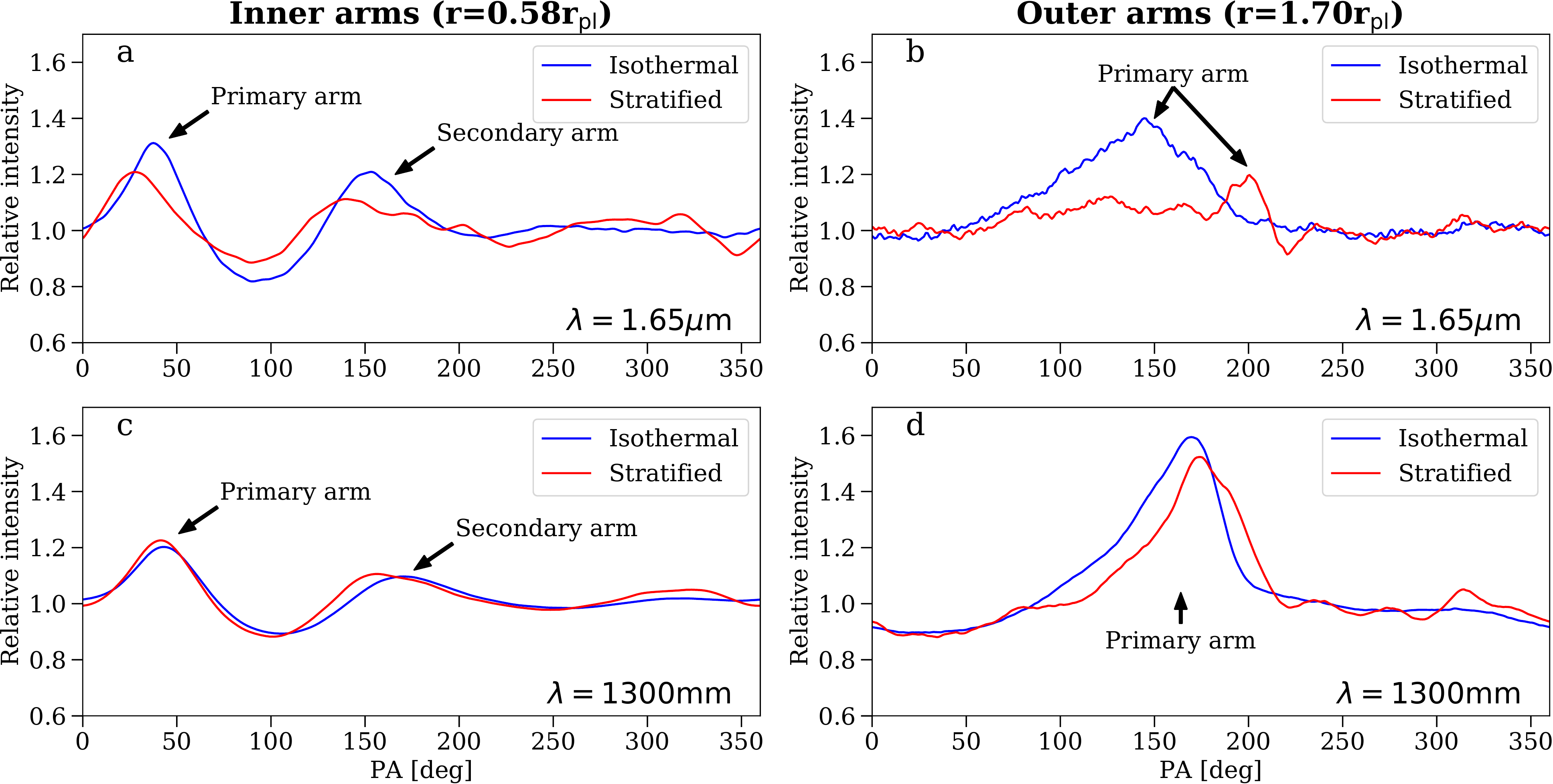}
\caption{Comparison of the azimuthal surface brightness profile of the disc in synthetic images calculated from locally isothermal and stratified 
simulations.  The comparison is shown for inner ({\it a}, {\it c}) and outer spirals ({\it b}, {\it d}) and both in NIR scattered light ({\it a}, {\it b})
and in the sub-millimetre continuum ({\it c}, {\it d}). The surface brightness profiles are normalised to the mean value in the [200$^\circ$, 300$^\circ$] 
interval. In the NIR the amplitude and the azimuthal width of the spirals are significantly larger in isothermal simulations compared to 
the stratified one, while the vertical structure has little effect on the sub-millimetre continuum observations. }
\label{fig:azim_profile}
\end{figure*}

We show the simulated NIR scattered light and sub-millimetre continuum images calculated from vertically isothermal (see Figure\,\ref{fig:images}a,b)
and thermally stratified (see Figure\,\ref{fig:images}{c,d}) hydrodynamic simulations. In terms of the general structure, the isothermal and stratified simulation look largely similar, with the only significant difference that the inner part of the disc in the stratified simulation is brighter with respect to the isothermal simulation. This is because the inner edge of the disc in the stratified simulation is more "rounded off", increasing the flaring index of the disc locally and consequently the brightness of the disc.

\subsubsection{Amplitude of the spirals}

In Figure\,\ref{fig:azim_profile} we compare the azimuthal surface brightness profiles of the two models at two radii, 1.7\,$r_p$ and 0.58\,$r_p$. From this plot we measure the peak-to-peak variation of the intensity in the azimuthal direction and the full width at half maximum of the spiral peak (for the inner spiral we considered the stronger, primary arm). Since the two models employ the same temperature in the disc midplane, in the sub-millimetre case (Figure\,\ref{fig:azim_profile}{\it b, d}) the peak-to-peak intensity variation and the width of the spiral peak agree within 10\,\%. In the near-infrared scattered light images (Figure\,\ref{fig:azim_profile}{\it a, c}) instead the spiral contrast is 
higher in the isothermal simulations by 49\% and 45\% in the inner and outer arms, respectively. This is caused by the higher temperature in the upper layers in the stratified simulations; the enhanced pressure forces tend to wash out more the spiral wake and reduce its amplitude. While the azimuthal width of the spirals 
in the two simulations agree to 4\% inwards of the planet, for the outer spirals in the isothermal models the azimuthal width is a factor 3.5 times larger compared to the stratified simulations.

\subsubsection{Spiral wake}

\begin{figure*}
\center
\includegraphics[width=15cm]{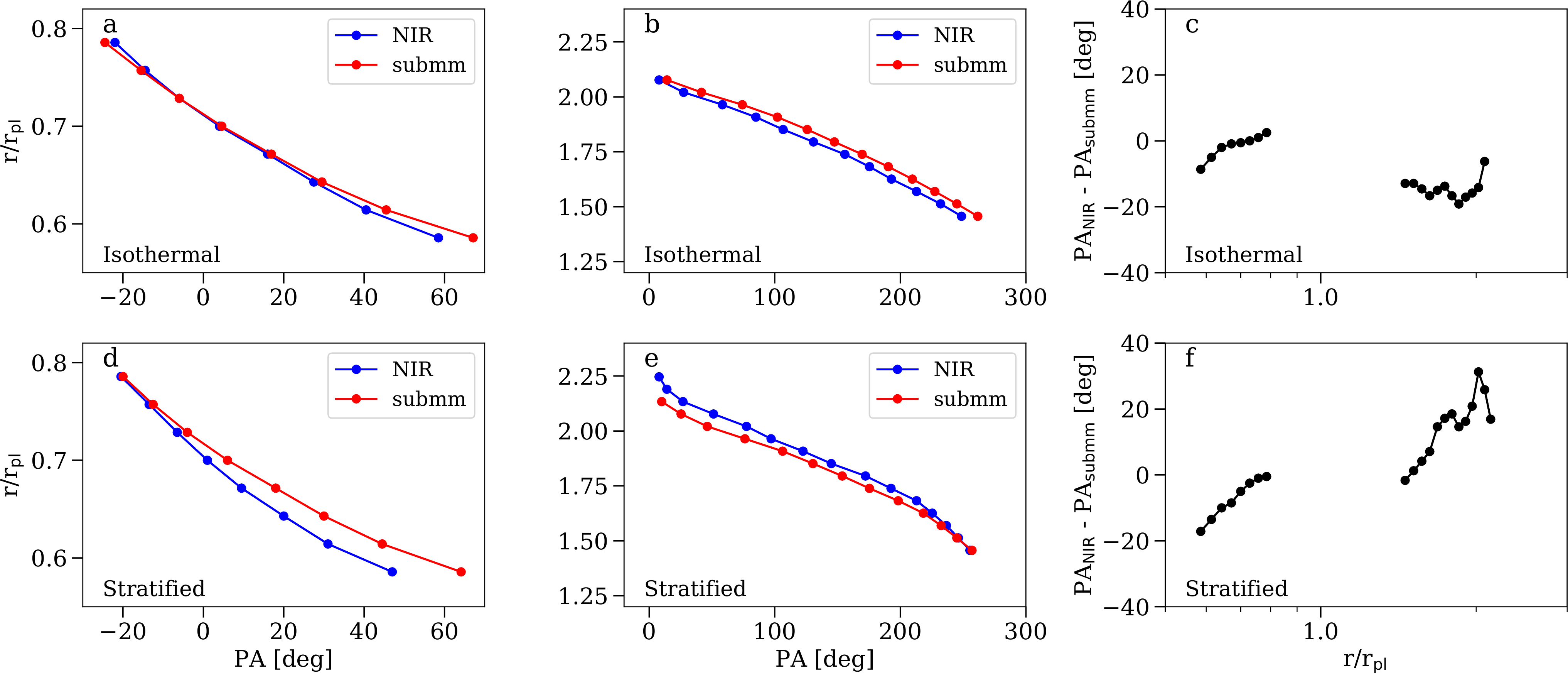}
\caption{Panel a and b show the spiral wake in the vertically isothermal simulation inwards and outwards of the planetary orbit, respectively. 
Panel c shows the difference in the azimuthal position of the spirals between sub-millimetre and NIR scattered light images as a function of radius.
Panel d--f are for the vertically stratified simulation. In a vertically isothermal disc the shape of the spiral wake
is largely the same at both wavelengths. \rev{For the stratified simulation the spirals are more open in the NIR than in the sub-millimetre both inwards and outwards of the planetary orbit.}
In the stratified simulation the difference in the azimuthal position of the spirals between NIR and sub-millimetre increases with increasing distance from the planet, i.e. the spirals converge towards the planet. 
}
\label{fig:wake}
\end{figure*}

In Figure\,\ref{fig:wake} we present the spiral wakes we derived from the synthetic images by extracting the positions of the local maxima in the radial surface 
brightness distribution. In sub-millimetre images, probing the disc mid-plane, the spirals have similar pitch angles in the locally isothermal and
in the vertically stratified simulations (see Figure\,\ref{fig:wake}{\it a} and Figure\,\ref{fig:wake}{\it d}) since the mid-plane temperatures in the two simulations (locally isothermal vs. stratified) are identical. 

Spirals in scattered light images computed from stratified simulations (see Figure\,\ref{fig:wake}{\it e}) instead are more open compared to the ones computed from locally isothermal simulations (see Figure\,\ref{fig:wake}{\it b}), consistently with the analysis of the simulation in section \ref{sec:res_sims}. \rev{For example, while we find that \textit{outer} spirals in NIR scattered light images have a low pitch angle (7.5\degr at 1.5$r_{\rm p}$) in the isothermal simulations, the pitch angle is higher in the stratified simulation (13\degr at 1.5$r_{\rm p}$); the effect is smaller for the inner spiral (15\degr vs 13\degr at 0.7$r_{\rm p}$). \citet{dong_2015a} suggested that a low pitch angle could be used to identify the spiral as an outer one, but it is not clear whether this method still applies to the stratified case.}

To better illustrate the difference in opening angle we show the 
difference in azimuth angles between the spirals derived from scattered light and sub-millimetre images as a function of stellocentric 
radius in Figure\,\ref{fig:wake}{\it c} and Figure\,\ref{fig:wake}{\it f} for the locally isothermal and for the stratified 
simulations, respectively. In the locally isothermal simulation the difference in the azimuthal position of the spiral wake is not larger than 8\degr
for the inner arms at all radii with a weak trend of increasing difference towards smaller radii. For the outer spirals the difference fluctuates 
between -19\degr and -6\degr with no clear dependence on the radius. 
In case of the stratified simulations there is a clear trend for the azimuthal position difference of the spirals in near-infrared and sub-millimetre. 
The azimuthal difference in the spiral position increases both inwards and outwards from the position of the planet. The difference in the spiral position in near-infrared and
sub-millimetre images is nearly -20\degr for the inner arms at 0.55$r_{\rm pl}$ in the stratified simulations compared to the -8\degr in the locally
isothermal simulations. For the outer arms the azimuthal position difference is about 30\degr at 2$r_{\rm pl}$ in the
stratified simulations while only -15\degr in the locally isothermal simulations. 

\section{Discussion and conclusions}
\label{sec:conclusions}

Direct detection of young planets is notoriously difficult due to the contrast limitation of the observations. Due to their larger spatial extent, spiral density waves excited by young planets are easier to detect. Thus, spiral arms could, in principle, be used as signposts of planets (see e.g. \citealt{dong_2015a}). However, there is ambiguity in whether the planet is \textit{inside} or \textit{outside} the spirals. In the previous section we have shown that the spiral arms in the sub-mm and in scattered light converge towards the location of the planet due to the different temperature in the midplane compared to the upper layers of the disc. Therefore, we suggest to use the direction of convergence of the spirals in near-infrared and in 
sub-millimetre continuum images to infer whether the planet is inwards or outwards of the spirals. 

\rev{Observationally, this requires having high-resolution observations of the same disc in near-infrared scattered light, probing the upper layers of the disc, and in sub-millimetre continuum, probing the disc mid-plane. Although not explored in this paper, we note that optically thin gas observations also trace the disc midplane and could therefore be used in place of the sub-mm continuum. 
}

We also note that, if we know that a planet is responsible for the perturbation, the dependence of the spiral pitch angle on the local temperature could be used as a diagnostic to map the vertical thermal stratification of the disc. 
This method requires using different tracers (e.g. NIR scattered light, various CO isotopologues, submm-continuum) coming from different heights above the midplane. Mapping the vertical thermal structure of the disc with the spiral pitch angle has the advantage over other alternative methods (e.g. modelling of the dust continuum emission, molecular rotational transitions), that the pitch angle depends only on the dynamics and is not affected by uncertainties in chemistry and opacities. While the \textit{absolute} value of the pitch angle might be  weakly affected by the planet mass or viscosity \citep{dong_2017}, the \textit{differential} variation of it as a function of height is likely independent of these factors. A detailed mapping will require a non-linear theory of spiral density waves in thermally stratified discs, which has yet to be developed.


\section*{Acknowledgements}
This work has been supported by the DISCSIM project, grant agreement 341137 funded by the European Research Council under ERC-2013-ADG. 
We thank the referee, Ruobing Dong, for the insightful review, that improved our paper.  

\bibliographystyle{mnras}
\bibliography{manuscript}




\label{lastpage}
\end{document}